\documentclass{aa}
\usepackage{graphicx,psfrag,amssymb,amsmath}
\usepackage{txfonts}
\usepackage{natbib}
\usepackage{epsfig}
\bibpunct{(}{)}{;}{a}{}{,}
\begin{document}
\title{The interpretation of water emission from dense interstellar clouds}
\author{Dieter R.~Poelman,
           \inst{1,2}
        Marco Spaans
           \inst{1}
           \and
        A.~G.~G.~M.~Tielens
           \inst{3,1}}
\offprints{D.R.Poelman@astro.rug.nl}
\institute{Kapteyn Astronomical Institute, P.O. Box 800, 9700 AV Groningen, the Netherlands\\
\email{D.R.Poelman@astro.rug.nl}
        \and
         SRON Netherlands Institute for Space Research, Landleven 12, 9747 AD Groningen, the Netherlands
        \and
         NASA Ames Research Center, MS245-3, Moffett Field, CA 94035, USA
           }

\date{Received / Accepted}
\abstract
{Existing SWAS observations and future HIFI/Herschel data require a clear sense of the information content of water emission and absorption lines.}
{Investigate wether the ground-state transition of ortho-$\mathrm{H_2O}$ ($\mathrm{1_{10}}$$\rightarrow$$\mathrm{1_{01}}$) at 557GHz can be used to measure the column density throughout an interstellar cloud.}
{We make use of a multi-zone escape probability code suitable for the treatment of molecular line emission.}
{For low abundances, i.e., X($\mathrm{H_2O}$) $\lesssim$ $10^{-9}$, the intensity of the $\mathrm{1_{10}}$$\rightarrow$$\mathrm{1_{01}}$ transition scales with the total column of $\mathrm{H_2}$. However, this relationship breaks down with increasing abundance, i.e., optical depth, due to line trapping and, -- for $\mathrm{T_{dust}}$$\gtrsim$25K, X($\mathrm{H_2O}$)$\lesssim$$10^{-8}$ and n$\sim$$10^4$$\mathrm{cm^{-3}}$, -- absorption of the dust continuum.}
{An observed decline in intensity per column, expected if $\mathrm{H_2O}$ is a surface tracer, does not necessarily mean that the water is absent in the gas phase at large columns, but can be caused by line trapping and subsequent collisional de-excitation. To determine the amount of water vapour in the interstellar medium, multiple line measurements of optically thin transitions are needed to disentangle radiative transfer and local excitation effects.}
\keywords{ISM: molecules  -- radiative transfer
          }
\authorrunning{Poelman et al.}
\titlerunning{The interpretation of water emission from dense interstellar clouds}
\maketitle

\section{\label{sec:intro}Introduction}
The launch of the {\em Submillimeter  Wave Astronomy Satellite} \citep[SWAS;][]{2000ApJ...539L..87M} made it possible to observe emission from the ground-state transition ($\mathrm{1_{10}}$$\rightarrow$$\mathrm{1_{01}}$) of ortho-$\mathrm{H_2^{16}O}$ and its isotopomer ortho-$\mathrm{H_2^{18}O}$, to determine the abundance of water vapour, i.e., column density $\mathrm{N_{o-H_2O}}$, in various regions in the interstellar medium, e.g., dense and diffuse interstellar gas clouds, circumstellar envelopes, planetary atmospheres, and comets \citep[e.g.,][]{2000ApJ...539L..87M, 2000ApJ...539L.101S, 2000ApJ...539L..93S, 1995Icar..117..162E, 2000ApJ...539L.147B, 2000ApJ...539L.143G, 2000ApJ...539L..87M}. In the future, the {\em Heterodyne Instrument for the Far Infrared} (HIFI) on {\em Herschel} will observe even more transitions of ortho- and para-$\mathrm{H_2O}$ (o/p-$\mathrm{H_2O}$) like $\mathrm{2_{12}}$$\to$$\mathrm{1_{01}}$ (1669.904 GHz), $\mathrm{2_{21}}$$\to$$\mathrm{2_{12}}$ (1661.015 GHz), $\mathrm{3_{03}}$$\to$$\mathrm{2_{12}}$ (1716.774 GHz), $\mathrm{3_{12}}$$\to$$\mathrm{3_{03}}$ (1097.357 GHz), $\mathrm{3_{21}}$$\to$$\mathrm{3_{12}}$ (1162.910 GHz), $\mathrm{1_{11}}$$\to$$\mathrm{0_{00}}$ (1113.342 GHz), $\mathrm{2_{02}}$$\to$$\mathrm{1_{11}}$ (967.924 GHz),  $\mathrm{2_{11}}$$\to$$\mathrm{2_{02}}$ (752.029 GHz) and  $\mathrm{2_{20}}$$\to$$\mathrm{2_{11}}$ (1228.801 GHz); some in absorption while others in emission.\\
One of the main goals of the SWAS mission is to determine where and whether $\mathrm{H_2O}$ and $\mathrm{O_2}$ are the major reservoirs of oxygen through the interstellar medium. SWAS observations have determined the gaseous water abundance in warm dense gas (T $\gtrsim$ 300 K and n($\mathrm{H_2}$) $\gtrsim$ $10^3$ \mbox{$\mathrm{cm}^{-3}$}) to be $10^{-5}$ relative to $\mathrm{H_2}$, in good agreement with chemical models for such conditions.
However, in cold (T $\lesssim$ 30 K), dense clouds the abundance of gaseous water is $\sim$100 to 1000 times below the predictions of cold-cloud gas-phase chemical models. It has been suggested that -- toward cold clouds -- gaseous $\mathrm{H_2O}$ exists only near the cloud surface. Indeed, closer to the surface than an $\mathrm{A_V}$ of a few mag, $\mathrm{H_2O}$ is photo-dissociated by the ambient galactic UV field. Deeper into the cloud, i.e., $\mathrm{A_V}$ of 4 $\sim$8 mag (depending on density and UV intensity), $\mathrm{H_2O}$ may rapidly deplete onto dust grains \citep{2000ApJ...539L.129B, 2001A&A...378.1024C, 2001A&A...370..557V}. Although the derivation of the column density from {\em absorption} observations is straightforward (column density is simply proportional to the optical depth in the line, see \cite{2004ApJ...605..247P}) this is not the case for {\em emission} observations. The analysis to determine the $\mathrm{H_2O}$ abundance now crucially depends on the physical properties of the gas through the collisional rate coefficients. Therefore, accurate constraints on the gas densities and temperatures are needed.\\
Extensive SWAS observations of the Orion A molecular cloud show that gaseous $\mathrm{H_2O}$ correlates with CN, a surface tracer, rather than with $\mathrm{C^{18}O}$, a volume tracer \citep{2005AdSpR..36.1027M}. This result has been interpreted as evidence that gaseous water resides only near the surface. However, caution is needed when relying purely on single transition observations to draw such a conclusion in view of the complex rotational level structure of the $\mathrm{H_2O}$ molecule. In particular, there is a fundamental difference between an optically thin and an effectively optically thin line. The latter case implies a strong coupling between line photons and water molecules. A full radiative transfer calculation is needed to address this problem, since the observed intensities of molecular emission depend on a complex competition between radiative and collisional processes. Moreover, the excitation of $\mathrm{H_2O}$ also differs from that of other molecules, since both collisions and infrared radiation from warm dust influence the level populations \citep{1983ApJ...275..145T}.\\
The intent of this work is to show that it is not straightforward to retrieve accurate information, e.g., column density, from single transition observations of $\mathrm{H_2O}$ due to the complex level structure of this molecule.

\section{\label{sec:}Basic model description}
The results presented here were obtained by application of the numerical code of \cite{2005A&A...440..559P}, described further in \cite{2006A&A...453..615P}. The interested reader is referred to these papers for a description of the underlying algorithms. The radiative transfer of o/p-$\mathrm{H_2O}$ is solved by means of a multi-zone escape probability method in three dimensions. By using a multi-zone formalism, the medium is divided into different zones, i.e., gridcells, each with a value for the abundance of the species (e.g., $\mathrm{H_2O}$), the density of the medium, and the temperature of gas and dust. Besides this, the cloud is characterized by a fixed total column density. The statistical equilibrium equation for a multilevel sytem can be written as
\begin{equation}
\sum_{i>j}(n_iA_{ij}~+~nn_i\gamma_{ij}~+~n_iB_{ij}J)~=~\sum_{i<j}(nn_i\gamma_{ij}~+~n_iB_{ij}J),
\label{eq:stateq}
\end{equation}
where the radiative de-excitation is given by the Einstein A coefficients, $\mathrm{B_{ij}}$ are the Einstein coefficients for absorption and stimulated emission, $\gamma_{ij}$ are the collisional rate coefficients, $J$ the mean photon radiation field, $n_i$ the population density of the $i$th level, and $n$ the density of the medium. 
The level populations depend on the radiation field while the radiation field depends on the level populations everywhere. 
By introducing the concept of a multi-zone escape probability \citep{2005A&A...440..559P, 2006A&A...453..615P} the statistical equilibrium equations are decoupled from the radiative transfer equations. Eq. \ref{eq:stateq} can be rewritten as
\begin{equation}
\sum_{i>j}(n_iA_{ij}\beta(\tau_{ij})~+~nn_i\gamma_{ij})~=~\sum_{i<j}(nn_i\gamma_{ij}),
\label{eq:levelpop}
\end{equation}
because the net absorptions, corrected for stimulated emission, are equal to those photons that do not escape, i.e.,
\begin{equation}
(n_jB_{ji}~-~n_iB_{ij})_{i>j}J~=~\sum_{i>j}n_i[1~-~\beta(\tau_{ij})]A_{ij},
\label{eq:approx}
\end{equation}
where $\beta(\tau)$ is the probability that a photon formed at optical depth $\tau$ in a certain direction escapes the cloud along that direction. Therefore, 
\begin{equation}
n_{cr}~=~\frac{\sum_{i>j}\beta(\tau_{ij})A_{ij}}{\sum_{i>j}\gamma_{ij}}
\label{Eq:ncr}
\end{equation}
Note that because of the large Einstein A coefficients of the water molecule critical densities are of the order of $10^8$-$10^9$ \mbox{$\mathrm{cm}^{-3}$} in the optically thin case.\\ 
The background radiation field P($\nu_{ij}$) in the code consists of two terms: the 2.7K microwave background and the infrared emission of dust at a temperature $T_{\mathrm{d}}$. This is,
\begin{equation}
P(\nu_{ij}) = B(\nu_{ij}, T=2.7K)\ +\ (1-e^{-{\tau_{dust}}})B(\nu_{ij},T_d)
\end{equation}
The intensity of transition i$\rightarrow$j (i$>$j) is then given by
\begin{equation}
I_{ij,\mathrm{total}} = {1\over 4\pi}\int_0^r\Lambda_{ij,\mathrm{local}}dr,
\label{eq:int_total}
\end{equation}
with 
\begin{equation}
\Lambda_{ij,\mathrm{local}} = n_iA_{ij}h\nu_{ij}\beta(\tau_{ij})\{[S(\nu_{ij})-P(\nu_{ij})]/S(\nu_{ij})\},
\end{equation} 
S($\nu_{ij}$) is the source function at frequency $\nu_{ij}$. \\
Collisional rate coefficients for inelastic collisions between o/p-$\mathrm{H_2O}$ and He \citep{1993ApJS...85..181G}, and for collisions between o/p-$\mathrm{H_2O}$ and both o-$\mathrm{H_2}$ and p-$\mathrm{H_2}$ \citep{1996ApJS..107..467P} are adopted. We adopt the expression for the ortho-to-para ratio (OPR) of $\mathrm{H_2}$, in thermal equilibrium, defined by 
\begin{equation}
\mathrm{OPR}= {{(2I_\mathrm{o} + 1)\sum(2J + 1)\exp\left(-{E_\mathrm{o}(J,K_\mathrm{a},K_\mathrm{c})\over kT}\right)}\over{(2I_\mathrm{p} + 1)\sum(2J + 1)\exp\left(-{E_\mathrm{p}(J,K_\mathrm{a},K_\mathrm{c})\over kT}\right)}}\,, 
\label{eq:OPR}
\end{equation}
where $I_\mathrm{o}$ and $I_\mathrm{p}$ are the total nuclear spin, corresponding to wether the hydrogen nuclear spins are parallel ($I_\mathrm{o}$ = 1, $\uparrow$$\uparrow$) or anti-parallel ($I_\mathrm{p}$ = 0, $\uparrow$$\downarrow$). The sum in the numerator (denominator) extends over all ortho (para) levels $({J},K_\mathrm{a},K_\mathrm{c})$, \citet{1987A&A...187..419M}. The code has been tested extensively against (analytical) benchmark problems presented at the radiative transfer workshop held in Leiden (2004), see \citet{2006A&A...453..615P}. It is found that the level populations are completely consistent with the solutions of other Monte Carlo and ALI codes, as presented in \citet{2005dmu..conf..431V}\\
\begin{table}
\begin{minipage}[b]{\columnwidth}
\renewcommand{\footnoterule}{}
\caption{Model parameters}
\label{tab:model}
\centering
\begin{tabular}{lcccccc}
\hline\hline  
Model & $\mathrm{n(H_2)}$ & size & ${\mathrm{X(H_2O)}^{a}}$ & ${\mathrm{N({H_2})}^{b}}$ & $\mathrm{T_{gas}}$ & $\mathrm{T_{dust}}$\\
 &  $[$\mbox{$\mathrm{cm}^{-3}$}$]$ & $[$\mbox{pc}$]$ & &$[$\mbox{$\mathrm{cm}^{-2}$}$]$ & $[$K$]$&$[$K$]$\\
\hline
I & $10^4$-$10^6$ & 0.002-0.2 & $10^{-10}$-$10^{-6}$ & 6$\times$$10^{21}$ & 50 & no \\
II & $10^4$-$10^6$ & 0.002-0.2 & $10^{-10}$-$10^{-6}$ & 6$\times$$10^{21}$ & 50 & 15\\
III & $10^4$-$10^6$ & 0.002-0.2 & $10^{-10}$-$10^{-6}$ & 6$\times$$10^{21}$ & 50 & 25\\
IV & $10^4$-$10^6$ & 0.002-0.2 & $10^{-10}$-$10^{-6}$ & 6$\times$$10^{21}$ & 50 & 50\\
V & $10^4$-$10^6$ & 0.002-0.2 & $10^{-10}$-$10^{-6}$ & 6$\times$$10^{21}$ & 30 & 15\\
\hline
\end{tabular}
\end{minipage}
$^{a}$ Abundance of $\mathrm{H_2O}$; $^{b}$ Total column density through the centre
\end{table}
\section{Model results}

\begin{figure}
\includegraphics[width=9cm]{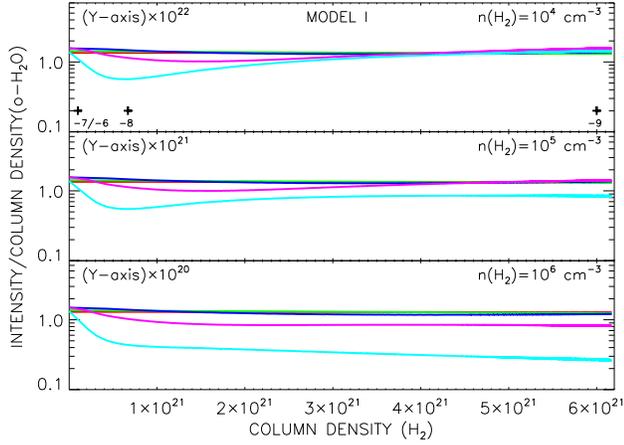}
\caption{The intensity of the ortho-$\mathrm{H_2O}$ ground state transition for a homogeneous sphere with $\mathrm{H_2}$ densities of $10^4$({\em top}), $10^5$({\em middle}), $10^6$({\em bottom}) $\mathrm{cm^{-3}}$ and temperature of the gas of 50 K. The dust emission as well as the CMB radiation are ignored ({\em i.e., model I}). In every case, the total column density ($\mathrm{H_2}$) is kept constant. Lines are plotted for an abundance of $\mathrm{H_2O}$, relative to $\mathrm{H_2}$, from $10^{-10}$ ({\em upper line/red}) to $10^{-6}$ ({\em lower line/light blue}) as function of $\mathrm{H_2}$ column density along the line of sight, where 2 $\times$ $10^{19}$ $\mathrm{cm^{-2}}$ is at the edge, and 6 $\times$ $10^{21}$ $\mathrm{cm^{-2}}$ through the center of the cloud. $Y$-axis in units of \mbox{$\mathrm{erg\ s^{-1}\ sr^{-1}}$}. The position of the $\tau$=1 surface is displayed with a cross for X($\mathrm{H_2O}$)=$10^{-9}$-$10^{-6}$.}
\label{fig:N6e21_nodustCMB}
\end{figure}

\begin{figure}
\includegraphics[width=9cm]{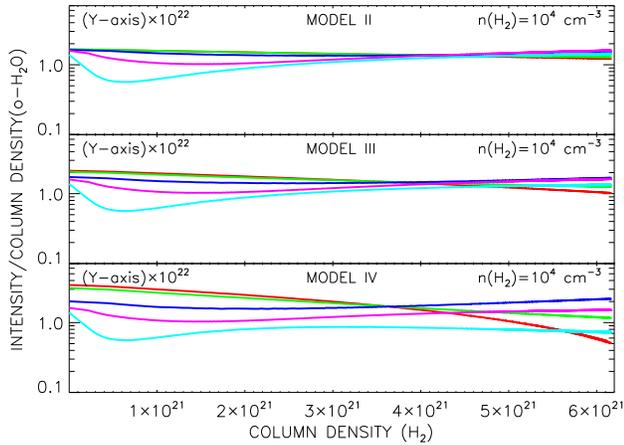}
\caption{The intensity of the ortho-$\mathrm{H_2O}$ ground state transition for a homogeneous sphere with density ($\mathrm{H_2}$) of $10^4$ $\mathrm{cm^{-3}}$ for model II ({\em top}), III ({\em middle}) and IV ({\em bottom}). Lines are plotted for an abundance of $\mathrm{H_2O}$, relative to $\mathrm{H_2}$, from $10^{-10}$ ({\em upper line/red}) to $10^{-6}$ ({\em lower line/light blue}) as function of $\mathrm{H_2}$ column density along the line of sight, where 2 $\times$ $10^{19}$ $\mathrm{cm^{-2}}$ is at the edge, and 6 $\times$ $10^{21}$ $\mathrm{cm^{-2}}$ through the center of the cloud. $Y$-axis in units of \mbox{$\mathrm{erg\ s^{-1}\ sr^{-1}}$}.}
\label{fig:N6e21_1e4.ps}
\end{figure}

\begin{figure}
\includegraphics[width=9cm]{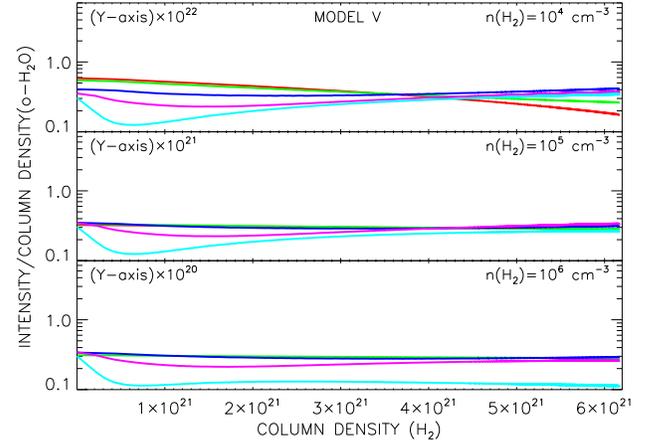}
\caption{The intensity of the ortho-$\mathrm{H_2O}$ ground state transition in case of a homogeneous sphere with densities ($\mathrm{H_2}$) of $10^4$({\em top}), $10^5$({\em middle}), $10^6$({\em bottom}) $\mathrm{cm^{-3}}$ and temperatures for gas and dust of 30 K and 15 K, respectively ({\em i.e., model V}). In every case, the total column density ($\mathrm{H_2}$) is kept constant. Lines are plotted for an abundance of $\mathrm{H_2O}$, relative to $\mathrm{H_2}$, from $10^{-10}$ ({\em upper line/red}) to $10^{-6}$ ({\em lower line/light blue}) as function of $\mathrm{H_2}$ column density along the line of sight, where 2 $\times$ $10^{19}$ $\mathrm{cm^{-2}}$ is at the edge, and 6 $\times$ $10^{21}$ $\mathrm{cm^{-2}}$ through the center of the cloud. $Y$-axis in units of \mbox{$\mathrm{erg\ s^{-1}\ sr^{-1}}$}.}
\label{fig:N6e21_Tg30Td15}
\end{figure}


\begin{figure}
\includegraphics[width=9cm]{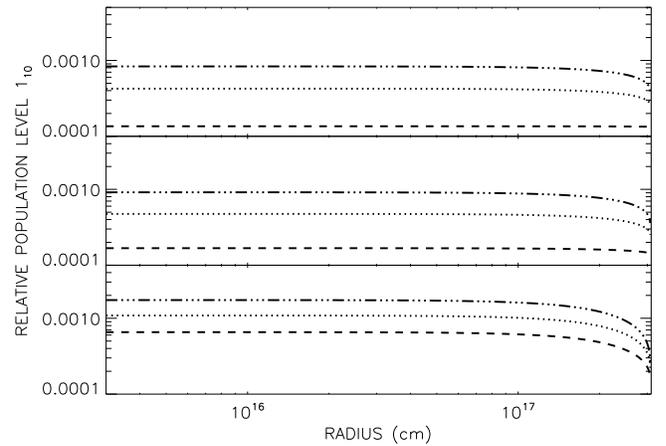}
\caption{Shown here is the population of the $\mathrm{1_{01}}$ level of o-$\mathrm{H_2O}$ as function of radius. Top, middle, bottom panel are the results in case X($\mathrm{H_2O}$) = $10^{-10}$, $10^{-9}$, $10^{-8}$, respectively. Dashed, dotted and dashed-dotted lines represent the results of model I, III and IV, respectively.}
\label{fig:levelpop}
\end{figure}

\begin{figure}
\includegraphics[width=9cm]{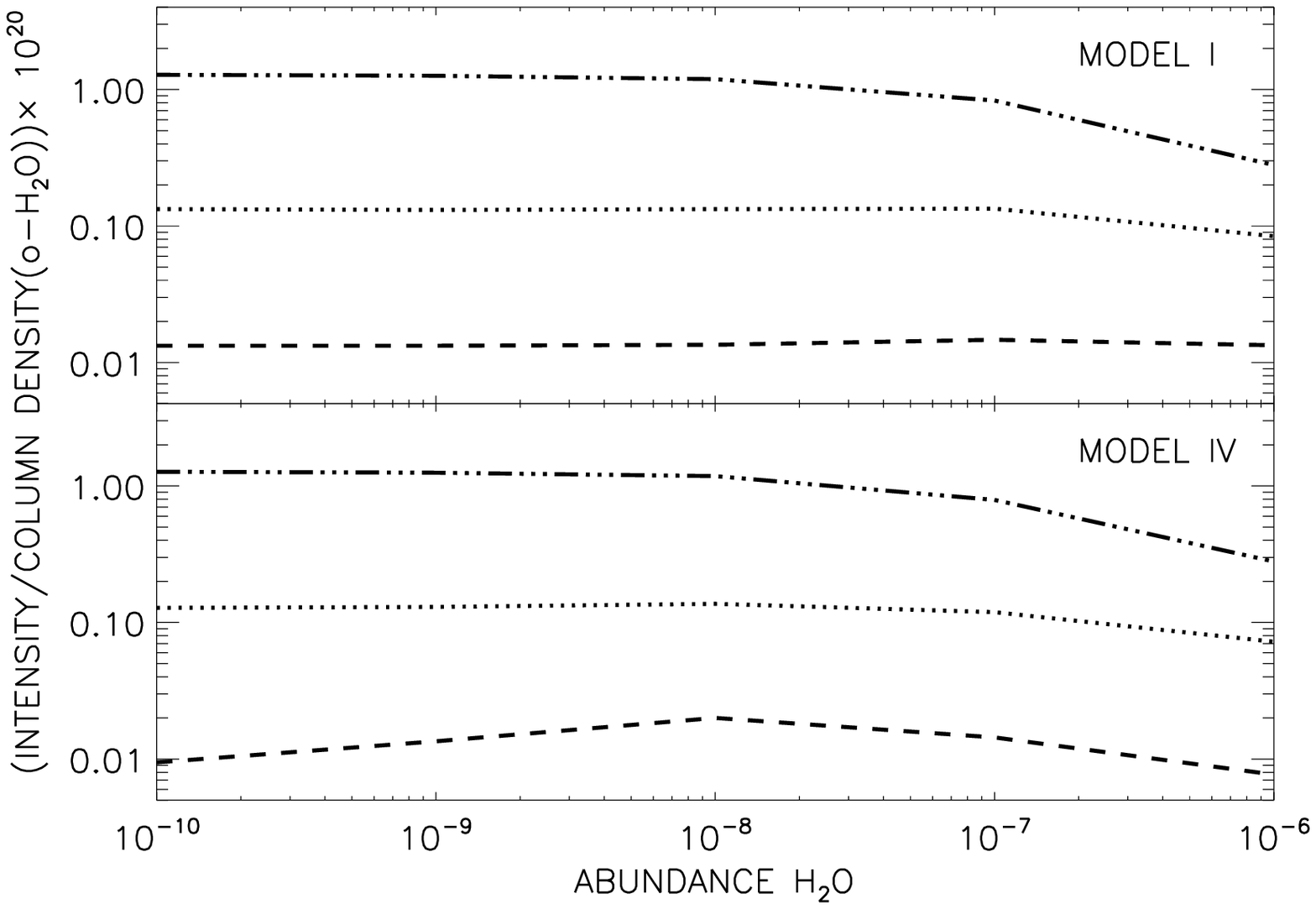} 
\caption{Conversion of the results of model I ({\em top}) and IV ({\em bottom}) into surface area weighted values. In each panel the dashed-dotted, dotted, and dashed curve represent the results in case the density n($\mathrm{H_2}$) is $10^6$, $10^5$ and $10^4$ \mbox{$\mathrm{cm}^{-3}$}, respectively. $Y$-axis in units of \mbox{$\mathrm{erg\ s^{-1}\ sr^{-1}}$}.}
\label{fig:mean}
\end{figure}
We calculate, in the case of a homogeneous sphere and as a function of impact parameter, the surface brightness for the $\mathrm{1_{10}}$$\rightarrow$$\mathrm{1_{01}}$ ground-state transition of ortho-$\mathrm{H_2O}$. The density and abundance of $\mathrm{H_2O}$ are the main parameters of the model. All the models have a constant total column density N($\mathrm{H_2}$) of 6 $\times$ $10^{21}$ \mbox{$\mathrm{cm}^{-2}$} through the center of the cloud, corresponding to a total $A_V$ of $\sim$3 mag. As a result, the physical size of the cloud is inversely proportional to the density of the medium. The density ranges from $10^4$ to $10^6$ \mbox{$\mathrm{cm}^{-3}$} covering the relevant range for dense molecular clouds  such as the Orion ridge. 
The temperatures, ranging from 30 to 50K for the gas and from 15 to 50K for the dust, were chosen to represent the mean observed temperatures towards star forming molecular clouds such as the Orion ridge which have been the focus of the SWAS effort. In model I we ignore the emission from dust and CMB, whereas in model IV a dust temperature of 50K is assumed, in order to assess the influence of the dust and temperature, respectively, on the excitation of the water molecule. The temperatures in model V are a factor of $\sim$2 lower than in the other models since to maintain a temperature of the gas of 50K throughout the cloud one needs a strong UV radiation field, which is not always the case. Note that the gas and dust temperatures are independent of cloud depth. The intent of the paper is to illustrate the excitation and radiative transfer effects assuming a 'simple' cloud model, not to model a realistic cloud. A Galactic dust-to-gas ratio of $10^{-2}$ by mass is assumed. We adopt the dust opacities of \cite{1994A&A...291..943O} (Column 5 of their Table 1). In all the models, except for model I, the dust optical depth $\tau_{\mathrm{dust}}$ through the center of the cloud at the frequency of the ground state transition of o-$\mathrm{H_2O}$, i.e., $\mathrm{\tau_{\nu=556.936GHz}}$ or $\mathrm{\tau_{\lambda=538\mu m}}$, is $10^{-3}$. Within each model the water abundance, X($\mathrm{H_2O}$), ranges from $10^{-10}$ to $10^{-6}$. The parameters for the different models are shown in Table \ref{tab:model}. Throughout the models, a velocity dispersion of 1 \mbox{$\mathrm{km\ s^{-1}}$} is adopted, typical for a cloud with moderate turbulence. Note that the results presented in this paper depend on the adopted velocity dispersion. A higher (lower) velocity dispersion will decrease (increase) the optical depth for a given transition, hence having an impact on the excitation of the molecule.\\
Fig. \ref{fig:N6e21_nodustCMB}--\ref{fig:N6e21_Tg30Td15} present the basic results of this work. We plot, as a function of N($\mathrm{H_2}$), i.e., different impact parameter, the $\mathrm{1_{10}}$$\rightarrow$$\mathrm{1_{01}}$ line intensity above the continuum per unit column density of o-$\mathrm{H_2O}$, see Eq. \ref{eq:int_total}. The results for model I, in which we ignore the contribution of dust and CMB on the excitation of the water molecule are displayed in Fig. \ref{fig:N6e21_nodustCMB}. The position of the $\tau$=1 surface is denoted by a cross for X($\mathrm{H_2O}$)=$10^{-9}$-$10^{-6}$. Models II, III and IV behave in a similar way as in model I in the case n($\mathrm{H_2}$)$\ge$$10^5$$\mathrm{cm^{-3}}$. For this reason, we only plot the outcome in the low density case for models II, III and IV in Fig. \ref{fig:N6e21_1e4.ps}. The results for model V are plotted in Fig. \ref{fig:N6e21_Tg30Td15}. \\
The following trends can be identified, which will be discussed in Sect. \ref{sec:discussion}. \\
First, in the case of n($\mathrm{H_2}$) $\ge$ $10^5$ $\mathrm{cm^{-3}}$ and X($\mathrm{H_2O}$) $\lesssim$ $10^{-8}$, i.e., $\tau$$<$10, a linear relationship holds between the number of photons escaping the cloud and the impact parameter, i.e., I/$\mathrm{N_{H_2O}}$ is constant. However, this relationship breaks down at high optical depth, i.e., $\mathrm{H_2O}$ $>$ $10^{-8}$, for all the models. Second, in all the models, except in model I, absorption occurs in the low density case when abundances are low, i.e., X($\mathrm{H_2O}$) $<$ $10^{-8}$, which becomes more apparent as the dust temperature increases. However, the amount of absorption is moderate as the self-reversal in the center of the line is small. Third, when the $\mathrm{H_2O}$ abundance exceeds $10^{-7}$, in all the models, the ratio of the intensity to the column density decreases near the edge of the cloud (N($\mathrm{H_2}$)$\sim$5$\times$$10^{20}$$\mathrm{cm^{-2}}$). Fourth, for high optical depth, i.e., X($\mathrm{H_2O}$)$\gtrsim$$10^{-7}$, and n($\mathrm{H_2}$)=$10^6$ $\mathrm{cm^{-3}}$, I/$\mathrm{N_{H_2O}}$ decreases with increasing column density. Fifth, lowering the gas and dust temperatures by a factor of $\sim$2 (model V) does not lead to significant differences in the shapes of the curves. That is, models II-IV and model V experience similar complicating radiative transfer effects.

\section{\label{sec:discussion}Discussion}
The asymmetry of the water molecule causes the rotational levels to split into a number of different ladders, so called 'K-ladders', characterized by different values of the projection of the angular momentum onto the principal axes of the molecule ($J_{K_-K_+}$). Radiative transitions occur rapidly between levels in each ladder but are much slower between levels in different ladders. This leads to a spectrum more complex than linear or symmetric top molecules, e.g., CO, $\mathrm{NH_3}$.
Hence, it is not straightforward to disentangle the different processes that contribute to the observed spectrum. We now describe the different effects that play a role in the interpretation of the figures. In this, the 'edge' and 'centre' of the cloud refers to an impact parameter of 1 and 0, respectively. Note that the use of spherical models leads to angular non-trivial re-distribution of line photons. The same holds for continuum photons within the line profile frequency range. \\ 
First, one can see in Figs. \ref{fig:N6e21_nodustCMB} and \ref{fig:N6e21_Tg30Td15} that the curves as function of total column density ($\mathrm{H_2}$) are constant for X($\mathrm{H_2O}$) $\lesssim$ $10^{-8}$ and n($\mathrm{H_2}$)$\ge$$10^5$$\mathrm{cm^{-3}}$. In this limit, collisional de-excitation and scattering effects are negligible. Eventually every photon produced in the cloud will escape the cloud with few interactions with the surrounding medium. Note that the number of scatterings $N$ to escape depends on the optical depth. In this regime $\tau$ $\ll$ 1, therefore few photons are scattered and $N$ $\approx$ $\tau$. With increasing optical depth, i.e., X($\mathrm{H_2O}$) $\gtrsim$ $10^{-8}$, more effects have to be taken into account. All models show a drop near the edge of the cloud. 
 Because of the increasing optical depth, line-scattering effects become important. Thus, line photons then tend to escape in the direction with the lowest optical depth rather than tangentially to the cloud surface, causing the dip near the edge. However, towards the centre of the cloud, the optical depth increases with orders of magnitude. The photons will undergo numerous scatterings for $\tau$ $\gg$ 1, with $N$ $\approx$ $\tau^2$, and eventually will escape in the line wings.\\
Second, 
at densities as low as $10^4$ $\mathrm{cm^{-3}}$, and abundances not exceeding $10^{-8}$ (modest optical depth), the line is strongly subthermally excited, and radiatively colder than the dust background. Hence, the line appears in absorption. The decrease in intensity/N($\mathrm{H_2O}$) shown in Fig. \ref{fig:N6e21_1e4.ps} (red and green curves) now indicates that lines of sight through the cloud center are no longer contributing evenly to the emissivity across their entire column. The line is not strictly in absorption yet, but it has developed an intensity dip around line center. Thus, the presence of dust causes the trend in the intensity per column in this regime to decrease \citep{1983ApJ...275..145T}. This behaviour is not seen in case n($\mathrm{H_2}$) $\gtrsim$ $10^5$ $\mathrm{cm^{-3}}$, as in this regime collisions are the dominant process in the excitation of the water molecule, thereby nullifying the effect of dust emission.
 The influence of dust on the excitation/level populations of water is plotted in Fig. \ref{fig:levelpop} where the relative population of the $\mathrm{1_{10}}$ level is displayed. One notices that for warmer dust level $\mathrm{1_{10}}$ is more populated. In essence, dust continuum emission will tend to drive the level populations towards a Boltzmann distribution at the temperature of the dust. For a given density, the effects of radiative excitation by dust continuum emission is more pronounced for higher dust temperatures (e.g., higher continuum intensities).\\
Third, the effect of photon trapping is to lower the density at which LTE is approached, i.e., after each absorption, the gas has a chance to collisionally de-excite the species and return the excitation energy to the thermal bath of the gas, see Eq. \ref{Eq:ncr}. For optically thin gas, the critical density of the ground state transition at 557 GHz of o-$\mathrm{H_2O}$ is $\sim$$10^8$ \mbox{$\mathrm{cm}^{-3}$} at 50 K. The optical depth, through the centre of the cloud, varies from 0.1 to $10^3$ when the abundance rises from $10^{-10}$ to $10^{-6}$ in all the models. Hence, for high abundances, i.e., X($\mathrm{H_2O}$) $\gtrsim$$10^{-7}$, the effective critical density drops to $10^5$-$10^6$ \mbox{$\mathrm{cm}^{-3}$}, since for high optical depth $\beta(\tau)$$\sim$$1/\tau$. Collisional de-excitation processes then become important in the regime where n($\mathrm{H_2}$) $\gtrsim$ $10^5$ \mbox{$\mathrm{cm}^{-3}$}, and X($\mathrm{H_2O}$) $\gtrsim$ $10^{-7}$. It is seen in Fig. \ref{fig:N6e21_nodustCMB} and \ref{fig:N6e21_Tg30Td15} that for n($\mathrm{H_2}$)=$10^6$$\mathrm{cm^{-3}}$ and X($\mathrm{H_2O}$)=$10^{-6}$ the I/$\mathrm{N_{H_2O}}$ drops as function of impact parameter. In this part of parameter space collisional de-excitation is important and the probability that line photons are lost to the thermal bath through collisional de-excitation during one of the many scattering events is high.\\
Fourth, calculations are performed for model V (Fig. \ref{fig:N6e21_Tg30Td15}) with gas temperatures a factor of 2 lower relative to the temperatures used in model II. We find that the shape of the curves are not affected by such a change in gas temperature. However, it affects the distribution of the level populations and thus the absolute intensity in the lines. Hence, temperature variations cannot dispense of the radiative transfer effects studied in this work.\\
To summarize, we plot in Fig. \ref{fig:mean}, as a function of abundance, the average intensity emanating from the cloud for model I and IV, i.e., 
\begin{equation}
{{\int I_{{1_{10}}\rightarrow{1_{01}}}(b)2\pi b\ db}\over{\int N_{\mathrm{H_2O}}(b)2\pi b\ db}}, 
\end{equation}
with b the impact parameter. One notices a drop by a factor of $\sim$2--5 in case n($\mathrm{H_2}$) is $10^5$--$10^6$ \mbox{$\mathrm{cm}^{-3}$}, respectively. Note that with the assumption of an effectively optically thin line one would underestimate the water column by these same factors.\\

\section{Astrophysical implications}
The intensity of the ground-state transition of o-$\mathrm{H_2O}$ is driven by a combination of the ambient gas and dust temperatures on the one side and by the density of the surrounding medium on the other side. It is this interplay, together with the complex structure of the molecule that drives the level populations. To interpret existing SWAS and future HIFI data a clear sense of the information content of the water 
 lines is needed. \\
SWAS observations of the lowest rotational transition of o-$\mathrm{H_2^{16}O}$ of the Orion A molecular cloud show that gaseous water correlates much better with the near surface tracer CN than with the volume tracer $\mathrm{C^{18}O}$, as presented in \citet{2005AdSpR..36.1027M}. Through these observations -- in which it is assumed that the ground-state transition of ortho-$\mathrm{H_2O}$ is effectively optically thin -- one concludes that water is a surface tracer. This is plausible from a chemical point of view in which photo-dissociation destroyes the water molecule near the surface. Further inwards the cloud the water abundance reaches its equilibrium value through photodesorption of $\mathrm{H_2O}$-ice and photodestruction of $\mathrm{H_2O}$-gas until it freezes-out onto dust grains deeper into the cloud. However, we have shown, as seen in Fig. \ref{fig:mean}, that for $\tau$$>$10 and n$>$$10^5$$\mathrm{cm^{-3}}$ the effectively optically thin assumption no longer holds. Hence, under these conditions one is limited to observing the '$\tau$=10' surface and cannot use water to trace the cloud's volume, even when it is present, i.e., not frozen out (Cernicharo, private communication). Therefore, as CN is a surface tracer and the water intensity originates from a layer of gas with an optical thickness of 1 -- depending on local excitation conditions this layer is a surface layer -- the CN intensity correlates much better with the $\mathrm{H_2O}$ intensity and not with the volume tracer $\mathrm{C^{18}O}$. Thus, the anti-correlation of $\mathrm{H_2O}$ with $\mathrm{C^{18}O}$ is partly due to optical depth effects, and is not neccesarily a result of chemical changes. As a consequence, the presence of water past the $\tau$=1 surface cannot be ruled out. \\
We would also like to point out here that the most interesting aspect of the correlation of the water line intensity with the CN line intensity is the fact that both are observed to vary by a factor $\sim$100. Theoretically, the CN abundance is expected to scale with density squared \citep{2005ApJ...632..302B}, indicating the importance of density variations over the Orion molecular cloud. Given the results presented in this paper, we surmise that these density variations will hamper the interpretation of the water observations.\\ 
In order to deduce the total water column along the line of sight, additional information is needed from other -- effectively optically thin-- lines, which will be observed with future missions such as Herschel/HIFI.

\begin{acknowledgements}
We are grateful to Ted Bergin and Gary Melnick for sending an early version of the manuscript. We also thank Floris van der Tak for helpful discussions and suggestions which have improved the paper and the anonymous referee for his/her constructive comments.
\end{acknowledgements}

\bibliographystyle{aa}
\bibliography{finalartikelIII}

\end{document}